\documentclass[journal, a4paper, technote]{IEEEtran}

\usepackage{graphicx}

\usepackage[cmex10]{amsmath}
\usepackage{mathrsfs}
\usepackage[tight,footnotesize]{subfigure}

\newtheorem{lem}{Lemma}
\newtheorem{thm}{Theorem}

\newtheorem{rem}{Remark}
\newtheorem{definition}{Definition}

\begin{document}
%
\title{On Unique Decodability}
%
%
%

\author{Marco~Dalai, Riccardo~Leonardi
\thanks{The authors are with the Department of Electronics for Automation,
        University of Brescia, via Branze 38 - 25123, Brescia, Italy.
        Email: \{marco.dalai, riccardo.leonardi\}@ing.unibs.it}
}
\maketitle

\begin{abstract}

In this paper we propose a revisitation of the topic of unique decodability and of some fundamental theorems of lossless coding.
It is widely believed that, for any discrete source $X$, every ``uniquely decodable'' block code satisfies
\begin{equation*}
E[l(X_1 X_2 \cdots X_n)]\geq H(X_1,X_2,\ldots,X_n),\nonumber
\end{equation*}
where $X_1, X_2,\ldots,X_n$ are the first $n$ symbols of the source, $E[l(X_1 X_2 \cdots X_n)]$ is the expected length of the code for those symbols and $H(X_1,X_2,\ldots,X_n)$ is their joint entropy.
We show that, for certain sources with memory, the above inequality only holds when a limiting definition of {``\em uniquely decodable code''} is considered. 
In particular, the above inequality is usually assumed to hold for any ``practical code'' due to a debatable application of McMillan's theorem to sources with memory. We thus propose a clarification of the topic, also providing an extended version of McMillan's theorem to be used for Markovian sources.

\end{abstract}



%
\IEEEpeerreviewmaketitle

\section{Introduction}

The problem of lossless encoding of information sources has been intensively studied over the years (see \cite[Sec. II]{verdu-tutorial} for a detailed historical overview of the key developments in this field). Shannon initiated the mathematical formulation of the problem in his major work \cite{shannon-1948} and provided the first results on the average number of bits per source symbol that must be used \emph{asymptotically} in order to represent an information source. 

For a random variable $X$ with alphabet $\mathcal{X}$ and probability mass function $p_X(\cdot)$, he defined the \emph{entropy} of $X$ as the quantity
\begin{equation*}
H(X)=\sum_{x\in\mathcal{X}} p_X(x)\log\frac{1}{p_X(x)}
\end{equation*}
On another hand, Shannon focused his attention on finite state Markov sources $X=\{X_1, X_2,\ldots\}$, for which he defined the \emph{entropy} as
\begin{equation*}
H(X)=\lim_{n\to \infty} \frac{1}{n}H(X_1,X_2,\ldots,X_n),
\end{equation*}
a quantity that is now usually called \emph{entropy rate} of the source.
Based on these definitions, he derived the fundamental results for fixed length and variable length codes. In particular, he showed that, by encoding sufficiently large blocks of symbols, the average number of bits per symbol used by fixed length codes can be made as close as desired to the entropy rate of the source while maintaining the probability of error as small as desirable. If variable length codes are allowed, furthermore, he showed that the probability of error can be reduced to zero without increasing the asymptotically achievable average rate. Shannon also proved the converse theorem for the case of fixed length codes, but he did not explicitly consider the converse theorem for variable length codes (see \cite[Sec. II.C]{verdu-tutorial}).

An important contribution in this direction came from McMillan \cite{mcmillan-kraftineq}, who showed  that every {\em ``uniquely decodable''} code using a $D$-ary alphabet must satisfy Kraft's inequality, $\sum_i D^{-l_i}\leq 1$, $l_i$ being the codeword lengths \cite{kraft-kraftineq}. Based on this result, he was able to prove that the expected length of a uniquely decodable code for a random variable $X$ is not smaller than its entropy, $E[l(X)]\leq H(X)$. 
This represents a strong converse result in coding theory. However, while the initial work by Shannon was explicitly referring to finite state Markov sources, McMillan's results basically considered only the encoding of a random variable. This leads to immediate conclusions on the problem of encoding memoryless sources, but an ad hoc study is necessary for the case of sources with memory. The application of McMillan's theorem to these type of sources can be found in \cite[Sec. 5.4]{cover-thomas-book} and \cite[Sec. 3.5]{gallager-book}. In these two well-known references, McMillan's result is used not only to derive a converse theorem on the asymptotic average number of bits per symbol needed to represent an information source, but also to deduce a non-asymptotic strong converse to the coding theorem. In particular, the famous result obtained (see \cite[Th. 3.5.2]{gallager-book}, \cite[Th. 5.4.2]{cover-thomas-book}, \cite[Sec. II, p. 2047]{role-of-pattern}) is that, for every source with memory, any uniquely decodable code satisfies
\begin{equation}\label{eq:intro_basicineq}
E[l(X_1 X_2 \cdots X_n)]\geq H(X_1,X_2,\ldots,X_n),
\end{equation}
where $X_1, X_2,\ldots,X_n$ are the first $n$ symbols of the source, $E[l(X_1 X_2 \cdots X_n)]$ is the expected length of the code for those symbols and $H(X_1,X_2,\ldots,X_n)$ represents their joint entropy.

In this paper we want to clarify that the above equation is only valid if a limiting definition of ``uniquely decodable code'' is assumed. In particular, we show that there are information sources for which a reversible encoding operation exists that produces a code for which equation \eqref{eq:intro_basicineq} does not hold any longer for every $n$. This is demonstrated through a simple example in Section \ref{sec:preview_ex}.
In Section \ref{sec:unique-decod} we revisit the topic of unique decodability,  consequently providing an extension of McMillan's theorem 
for the case of first order Markov sources. Finally, in Section \ref{sec:remarks}, some additional interesting remarks on the considered topic are made.

\section{A Meaningful Example}

\label{sec:preview_ex}
\begin{figure}
\centering
\includegraphics[width=4.5cm]{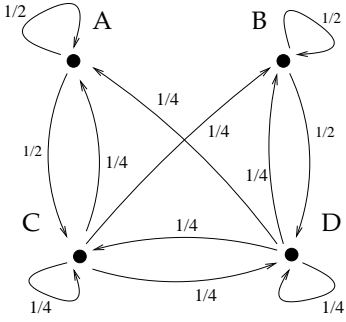}
\caption{Graph, with transition probabilities, for the Markov source use in the example.}
\label{fig:markov_prevexample}
\end{figure}

Let $X=\{X_1,X_2,\ldots\}$ be a first order Markov source with alphabet $\mathcal{X}=\{A,B,C,D\}$ and with transition probabilities shown by the graph of Fig. \ref{fig:markov_prevexample}. Its transition probability matrix is thus
\begin{equation*}
\mathbf{P}=\left[
\begin{array}{cccc}
1/2 & 0 & 1/2 & 0\\
0 & 1/2 & 0 & 1/2\\
1/4 & 1/4 & 1/4 & 1/4\\
1/4 & 1/4 & 1/4 & 1/4
\end{array}
\right],
\end{equation*}
where rows and columns are associated to the natural alphabetical order of the symbol values $A, B, C$ and $D$.

It is not difficult to verify that the stationary distribution associated with this transition probability matrix is the uniform distribution. Let $X_1$ be uniformly distributed, so that the source $X$ is stationary and, in addition, ergodic.

Let us now examine possible binary encoding techniques for this source and possibly find an optimal one. In order to evaluate the performance of different codes we determine the entropy of the sequences of symbols that can be produced by this source. By stationarity of the source, one easily proves that 
\begin{eqnarray}
H(X_1,X_2,\ldots,X_n) &= & H(X_1)+\sum_{i=2}^n H(X_i|X_{i-1})\nonumber\\
 & = & 2+\frac{3}{2}(n-1),\nonumber
\end{eqnarray}
where $H(X_i|X_{i-1})$ is the conditional entropy of $X_i$ given $X_{i-1}$, that is \begin{equation*}
H(X_i|X_{i-1}) = \\ \sum_{x,y\in \mathcal{X}}p_{X_iX_{i-1}}(x,y)\log \frac{1}{p_{X_i|X_{i-1}}(x|y)}.
\end{equation*}

Let us now consider the following binary codes to represent sequences produced by this source.

\vspace{0.3cm}
\noindent\textbf{Classic code}

We call this first code ``classic'' as it is the most natural way to encode the source given its particular structure.
Since the first symbol is uniformly distributed between four choices, 2 bits are used to uniquely identify it, in an obvious way. For the next symbols we note that we always have dyadic conditional probabilities. So, we apply a state-dependent code. For encoding the $k$-th symbol we use, again in an obvious way, 1 bit if symbol $k-1$ was an $A$ or a $B$, and we use 2 bits if symbol $k-1$ was a $C$ or a $D$. This code seems to perfectly fulfill the source as the number of used bits always corresponds to the uncertainty. Indeed, the average length of the code for the first $n$ symbols is given by
\begin{eqnarray*}
E[l(X_1,X_2,\ldots,X_n)] & \stackrel{}{=} & E[l(X_1)]+\sum_{i=2}^n E[l(X_i)]\\
 & =  & 2+\frac{3}{2}(n-1).
\end{eqnarray*}
So, the expected number of bits used for the first $n$ symbols is exactly the same as their entropy, which would let us declare that this encoding technique is optimal.

\vspace{0.3cm}
\noindent\textbf{Alternative code}

Let us consider a different code, obtained by applying the following fixed mapping from symbols to bits: $A\rightarrow 0$, $B\rightarrow 1$, $C\rightarrow 01$, $D\rightarrow 10$. 
It will be easy to see that this code maps different sequences of symbols into the same codeword. For example, the sequences $AB$ and $C$ are both coded to $01$. This is usually expressed, see for example \cite{cover-thomas-book}, by saying that the code is not \emph{uniquely decodable}, an expression which suggests the idea that the code cannot be inverted, different sequences being associated to the same code. It is however easy to notice that, for the source considered in this example, the code does not introduce any ambiguity. Different sequences that are producible by the source are in fact mapped into different codes. Thus it is possible to ``decode'' any sequence of bits without ambiguity. For example the code $01$ can only be produced by the single symbol $C$ and not by the sequence $AB$, since our source cannot produce such sequence (the transition from $A$ to $B$ being impossible).
It is not difficult to verify that it is indeed possible to decode any sequence of bits by operating in the following way. Consider first the case when there are still two or more bits to decode. In such a case, for the first pair of encountered bits, if a $00$ (respectively a $11$) is observed then clearly this corresponds to an $A$ symbol followed by a code starting with a 0 (respectively a $B$ symbol followed by a code starting with a 1). If, instead, a $01$ pair is observed (respectively a $10$) then a $C$ must be decoded (respectively a $D$). Finally, if there is only one bit left to decode, say a 0 or a 1, the decoded symbol is respectively an $A$ or a $B$.
Such coding and decoding operations are summarized in Table \ref{tab:altern_code}.
\begin{table}[b]
\begin{equation*}
\begin{array}{|l|c|}
\hline
\mbox{\bf Encoding} & 
\begin{array}{ccl}
A & \to & 0\\
B & \to & 1\\
C & \to & 01\\
D & \to & 10\\
\end{array}\\
\hline
\hline
\mbox{\bf Decoding} & 
\begin{array}{c|c}
\multicolumn{2}{c}{} \\
\mbox{more bits left} & \mbox{one bit left} \\
\hline\\
\begin{array}{ccl}
00\ldots & \to & A+0\ldots\\
01\ldots & \to & C\ldots\\
10\ldots & \to & D\ldots\\
11\ldots & \to & B+1\ldots\\
\end{array}
&
\begin{array}{ccl}
0 & \to & A\\
1 & \to & B\\
\\
\\
\end{array}
\end{array}\\
\hline
\end{array}
\end{equation*}
\caption{Table of encoding and decoding operations of the proposed alternative code for the Markov source of Figure \ref{fig:markov_prevexample}.}
\label{tab:altern_code}
\end{table}

Now, what is the performance of this code?
The expected number of bits in coding the first $n$ symbols is given by:
\begin{eqnarray*}
E[l(X_1 X_2 X_3 \cdots X_n)] & = & \sum_{i=1}^n E[l(X_i)]\\
 & = & \frac{3}{2}n\nonumber
\end{eqnarray*}
Unexpectedly, the average number of bits used by the code is strictly smaller than the entropy of the symbols. So, the performance of this code is better than what would have been traditionally considered the ``optimal'' code, that is the classical code. Let us mention that this code is not only more efficient on average, but it is at least as efficient as the classic code for every possible  sequence which remains compliant with the source characteristics.
For each source sequence, indeed, the number of decoded symbols after reading the first $m$ bits of the alternative code is always larger than or equal to the number of symbols decoded with the first $m$ bits of the classic code. Hence, the proposed alternative code is more efficient than the classic code in all respects. The obtained gain \emph{per symbol} obviously goes to zero asymptotically, as imposed by the Asymptotic Equipartition Property. However, in practical cases we are usually interested in coding a finite number of symbols. Thus, this simple example reveals that the problem of finding an optimal code is not yet well understood for the case of sources with memory. 
The obtained results may thus have interesting consequences not only from a theoretical point of view, but even for practical purposes in the case of sources exibiting constraints imposing high order dependencies.

Commenting on the ``alternative code'', one may object that it is not fair to use the knowledge on impossible transitions in order to design the code. But probably nobody would object to the design of what we called the ``classic code''. Even in that case, however, the knowledge that some transitions are impossible was used, in order to construct a state-dependent ``optimal'' code. 
\vspace{0.3cm}

It is important to point out that we have just shown a fixed to variable length code for a stationary ergodic source that maps sequences of $n$ symbols into strings of bits that can be decoded and such that \textbf{the average code length is smaller than the entropy of those $n$ symbols}. Furthermore, this holds for every $n$, and not for an {\em a priori} fixed $n$.
In a sense we could say that the given code has a negative \emph{redundancy}. 
Note that there is a huge difference between the considered setting and that of the so called \emph{one-to-one codes} (see for example \cite{Alon-onetoone} for details). In the case of one-to-one codes, it is assumed that only one symbol, or a given known amount of symbols, must be coded, and codes are studied as maps from symbols to binary strings without considering the decodability of concatenation of codewords. Under those hypotheses, Wyner \cite{Wyner-onetoone} first pointed out that the average codeword length can always be made lower than the entropy, and different authors have studied bounds on the expected code length over the years \cite{Blundo-one-to-one, Savari-onetoone}. Here, instead, we have considered a fixed-to-variable length code used to compress sequences of symbols of whatever length, concatenating the code for the symbols one by one, as in the most classic scenario.

\section{Unique decodability for constrained sources}

\label{sec:unique-decod}
In this section we briefly survey the literature on unique decodability and we then propose an adequate treatment of the particular case of \emph{constrained sources} defined as follows.
\begin{definition}
A source $X=\{X_1, X_2,\ldots\}$ with symbols in a discrete alphabet $\mathcal{X}$ is a \emph{constrained source} if there exists a finite sequence of symbols from $\mathcal{X}$ that cannot be obtained as output of the source $X$.
\end{definition}

\subsection{Classic definitions and revisitation}

It is interesting to consider how the topic of unique decodability has been historically dealt with in the literature and how the results on unique decodability are used to deduce results on the expected length of codes. Taking \cite{gallager-book} and \cite{cover-thomas-book} as representative references for what can be viewed as the classic approach to lossless source coding, we note some common structures between them in the development of the theory, but also some interesting differences.
The most important fact to be noticed is the use, in both references with only marginal differences, of the following chain of deductions:
\begin{enumerate}
\renewcommand{\labelenumi}{(\alph{enumi})}
\renewcommand{\theenumi}{(\alph{enumi})}

\item \label{deduct:wrong}McMillan's theorem asserts that all uniquely decodable codes satisfy Kraft's inequality;
\item If a code for a random variable $X$ satisfies Kraft's inequality, then $E[l(X)]\geq H(X)$;
\item Thus any uniquely decodable code for a random variable $X$ satisfies $E[l(X)]\geq H(X)$;
\item \label{deduct:last}For sources with memory, by considering sequences of $n$ symbols as {\em super-symbols}, we deduce that any uniquely decodable code satisfies $E[l(X_1, X_2,\ldots,X_n)]\geq H(X_1, X_2,\ldots,X_n)$.
\end{enumerate}

In the above flow of deductions there is an implicit assumption which is not obvious and, in a certain way, not clearly supported. It is implicitly assumed that the definition of \emph{uniquely decodable code} used in McMillan's theorem is also appropriate for sources with memory. Of course, by  definition of ``definition'', one can freely choose to define ``uniquely decodable code'' in any preferred way. However, as shown by the code of Table \ref{tab:altern_code} in the previous section, the definition of \emph{uniquely decodable code} used in McMillan's theorem does not coincide with the intuitive idea of ``decodable'' for certain sources with memory. To our knowledge, this ambiguity has never been reported previously in the literature, and for this reason it has been  erroneously believed that the result $E[l(X_1, X_2,\ldots,X_n)]\geq H(X_1, X_2,\ldots,X_n)$ holds for every ``practically usable'' code. As shown by the Markov source example presented, this interpretation is incorrect.

In order to better understand the confusion associated to the meaning of ``uniquely decodable code'', it is interesting to focus on a small difference between the formal definitions given by the authors in \cite{cover-thomas-book} and in \cite{gallager-book}. We start by rephrasing for notational convenience the definition given by Cover and Thomas in \cite{cover-thomas-book}. \begin{quote}
\begin{definition}{\cite[Sec. 5.1, pp. 79-80]{cover-thomas-book} }
\label{def:cover}
A code is said to be uniquely decodable if no finite sequence of code symbols can be obtained in two or more different ways as a concatenation of codewords.
\end{definition}
\end{quote}
Note that this definition is the same used in McMillan's paper \cite{mcmillan-kraftineq}, and it considers a property of the codebook without any reference to sources. It is however difficult to find a clear motivation for such a source independent definition. After all, a code is always designed for a given source, not for a given alphabet. Indeed, right after giving the formal definitions, the authors comment 
\begin{quote}
{\emph{``In other words, any encoded string in a uniquely decodable code has 
only one possible \emph{source string} producing it.''}}
\end{quote}
So, a reference to sources is introduced. What is not noticed is that the condition given in the formal definition coincides with the phrased one only if the source at hand can produce any possible combination of symbols as output. Conversely, the two definitions are not equivalent, the first one being stronger, the second one being instead ``more intuitive''.

With respect to formal definitions, Gallager proceeds in a different way with the following:
\begin{quote}
\begin{definition}{\cite[Sec. 3.2, pg. 45]{gallager-book}}
\label{def:gallager}
\emph{``A code is uniquely decodable if for each source sequence of finite length, the sequence of code letters corresponding to that source sequence is different from the sequence of code letters corresponding to any other source sequence.''}
\end{definition}
\end{quote}
Note that this is a formal definition of unique decodability of a code with respect to a given source. Gallager states this definition while discussing memoryless sources\footnote{See \cite[pg. 45]{gallager-book} \emph{``We also assume, initially, [...] that successive letters are independent''}}. In that case, the definition is clearly equivalent to Definition \ref{def:cover} but, unfortunately, Gallager implicitly uses Definition \ref{def:cover} instead of Definition \ref{def:gallager} when dealing with sources with memory.\footnote{In fact, in \cite{gallager-book}, the proof of Theorem 3.5.2, on page 58, is based on Theorem 3.3.1, on page 50, the proof of which states: \emph{``...follows from Kraft's inequality, [...] which is valid for any uniquely decodable code''}. But Kraft's inequality is valid for uniquely decodable codes defined as in Definition \ref{def:cover} and not Definition \ref{def:gallager}.}

In order to avoid the above discussed ambiguity, we propose to adopt the following explicit definition.
\begin{definition}
A code $C$ is said to be \emph{uniquely decodable for the source} $X$ if no two different finite sequences of source symbols producible by $X$ have the same code. 
\label{def:unique-decode-4source}
\end{definition}

With this definition, not all \emph{uniquely decodable codes for a given source} satisfy Kraft's inequality. So, the chain of deductions \ref{deduct:wrong}-\ref{deduct:last} listed at the beginning of this section cannot be used for constrained sources, as McMillan's theorem uses Definition \ref{def:cover} of unique decodability.

The alternative code of Table \ref{tab:altern_code} thus immediately gives:
\begin{lem}
There exists at least one source $X$ and a uniquely decodable code for $X$ such that, for every $n\geq 1$,
\begin{equation*}
E[l(X_1,X_2,\ldots,X_n)]<H(X_1,X_2,\ldots,X_n).
\end{equation*}
\end{lem}

\subsection{Extension of McMillan's theorem to Markov sources}

In Section \ref{sec:preview_ex}, the proposed alternative code demonstrates that McMillan's theorem does not apply in general to uniquely decodable codes for a constrained source $X$ as defined in Definition \ref{def:unique-decode-4source}. In this section a modified version of Kraft's inequality is proposed which represents a necessary condition for the unique decodability of a code for a first order Markov source.

Let $X$ be a Markov source with alphabet $\mathcal{X}=\{1, 2, \ldots, m\}$ and transition probability matrix $\mathbf{P}$. Let $W=\{w_1, w_2, \ldots, w_m\}$ be a set of $D$-ary codewords for the alphabet $\mathcal{X}$ and let, $l_i=l(w_i)$ be the length of codeword $w_i$.
McMillan's original theorem can be stated in the following way:

\begin{thm}[McMillan, \cite{mcmillan-kraftineq}]
\label{theo:kraft}
If the set of codewords $W$ is uniquely decodable (in the sense of Definition \ref{def:cover}) then
\begin{equation*}\label{eq:kraftineq}
\sum_{i=1}^m D^{-l_i}\leq 1.
\end{equation*}
\end{thm}

We propose a modified theorem for considering the unique decodability for the specific source.

\begin{thm}\label{theo:modified-kraft}
If the set of codewords $W$ is uniquely decodable for the Markov source $X$, then the matrix $\mathbf{Q}$ defined by
\begin{equation*}
\mathbf{Q}_{ij}=
\begin{cases}
0 & \mbox{if} \quad {P}_{ij}=0\\
D^{-l_j} & \mbox{if}\quad  {P}_{ij}>0
\end{cases}
\end{equation*}
has spectral radius at most 1.
\end{thm}
\begin{proof}
The proof is very similar to Karush's proof of McMillan's theorem \cite{karush-1961}.
Let $\mathcal{X}^{(k)}$ be the set of all sequences of $k$ symbols that can be produced by the source and let $\mathbf{L}=[D^{-l_1},D^{-l_2},\ldots,D^{-l_m}]'$. 
For $k>0$, define the row vector
\begin{equation*}
\mathbf{V}^{(k)}=\mathbf{L}'{\mathbf{Q}}^{k-1}.
\end{equation*}
It is easy to see by induction that the $i$-th component of $\mathbf{V}^{(k)}$ can be written as
\begin{equation*}
\mathbf{V}^{(k)}_i=\sum_{h_1, h_2, \ldots, h_k}D^{-l_{h_1}-l_{h_2}\cdots -l_{h_k}}
\end{equation*}
where the sum runs over all sequences of indices $(h_1, h_2,\ldots, h_k)$ with varying $h_1,h_2,\ldots,h_{k-1}$ and $h_k=i$ such that $(h_1,h_2,\ldots, h_k)\in \mathcal{X}^{(k)}$.
So, calling $\mathbf{1_m}$ the vector composed of $m$ 1's, we have
\begin{equation*}
\mathbf{L}'{\mathbf{Q}}^{k-1}\mathbf{1}_m =\sum_{(h_1, h_2, \ldots, h_k)\in\mathcal{X}^{(k)} }D^{-l_{h_1}-l_{h_2}\cdots -l_{h_k}}.
\end{equation*}
Reindexing the sum with respect to the total length $r=l_{h_1}+l_{h_2}+\cdots+l_{h_k}$ and calling $N(r)$ the number of sequences of $\mathcal{X}^{(k)}$ which are mapped in a length $r$ code, we have
\begin{equation*}
\mathbf{L}'{\mathbf{Q}}^{k-1}\mathbf{1}_m=\sum_{r=1}^{k l_{\text{max}}}N(r) D^{-r}
\end{equation*}
where $l_{\text{max}}$ is the maximum of the values $l_i,i=1,2,\ldots,m$.
Since the code is uniquely decodable for the source $X$, there are at most $D^r$ source-compatible sequences with a code of length $r$, that is, $N(r)\leq D^r$. Hence, for every $k>0$ 
\begin{equation}\label{eq:disug_powerlinear}
\mathbf{L}'{\mathbf{Q}}^{k-1}\mathbf{1}_m \leq \sum_{r=1}^{k l_{\text{max}}}D^{r} D^{-r}=k l_{\text{max}}
\end{equation}
Now, note that the irreducible matrix $\mathbf{Q}$ is also nonnegative. Thus, by the Perron-Frobenius theorem (see \cite{minc-book} for details), its spectral radius $\rho(\mathbf{Q})$ is also an eigenvalue, with algebraic multiplicity 1 and with positive associated left eigenvector. 

Suppose now $\rho(\mathbf{Q})>1$. Since $\mathbf{L}$ and $\mathbf{1}_m$ are both positive, it is easy to deduce that the term on the left hand side of equation (\ref{eq:disug_powerlinear}) asymptotically grows as $\rho(\mathbf{Q})^{k-1}$ when $k$ goes to infinity. On the contrary, the right hand side term only grows linearly with $k$ and, for large enough $k$, equation (\ref{eq:disug_powerlinear}) cannot hold. We conclude that $\rho(\mathbf{Q})\leq 1$.
\end{proof}

\section{Some Additional Remarks}
\label{sec:remarks}

\begin{rem}[Theorem \ref{theo:modified-kraft} generalizes Theorem \ref{theo:kraft}]
In the case of unconstrained Markov sources, Theorem \ref{theo:modified-kraft} is equivalent to Theorem \ref{theo:kraft}. Indeed, the Markov source being not constrained means that its transition probability matrix $\mathbf{P}$ has all strictly positive entries. This implies that the matrix $\mathbf{Q}$ defined in Theorem \ref{theo:modified-kraft} has all equal rows. The spectral radius of such a matrix equals the sum of the elements in every row, which is $\sum_{j}D^{-l_j}$, reducing thus to the classic Kraft's inequality.
\end{rem}

\begin{rem}[Non sufficiency of the condition]
Kraft's inequality is both a necessary and sufficient condition for the existence of a uniquely decodable code (in the sense of Definition \ref{def:cover}) with codeword lengths $l_i$. Theorem \ref{theo:modified-kraft}, instead, only gives a necessary condition on the lengths $l_i$ for the unique decodability of a code for a given source. It is easy to show that condition stated in the theorem is not a sufficient condition for the existence of a uniquely decodable code for a source with codeword lengths $l_i$. Finding a necessary and sufficient condition seems to be a much harder problem.
\end{rem}

\begin{rem}[Extended Sardinas-Patterson test]
With respect to the previous remark, we point out that it is however possible to test a given code for decodability for a given source by devising a generalization of the Sardinas-Patterson test \cite{sardinas-patterson-1953} to deal with constrained sources (see  \cite{DL_ITreport_2008}).
\end{rem}

\begin{rem}[A more general form of Theorem \ref{theo:modified-kraft}]
Theorem \ref{theo:modified-kraft} was formulated for the case of Markov chains ``in the Moore form'', as considered for example in \cite{cover-thomas-book}. In other words, we have modeled information sources as Markov chains by assigning an output source symbol to every state. In order to deal with more general sources we can consider Markov sources in the ``Mealy form'', where output symbols are not associated to states but to transitions between states (which corresponds to the Markov source model used by Shannon in \cite{shannon-1948} or, for example, by Gallager in \cite{gallager-book}). 
Theorem \ref{theo:modified-kraft} can be extended to this type of Markov sources as follows (see  \cite{DL_ITreport_2008}).

\begin{thm}\label{theo:modified-kraft-mealyform}
Let $X$ be a finite state source, with possible states $S_1, S_2, 	\ldots, S_q$ and with output symbols in the alphabet $\mathcal{X}=\{1,2,\ldots,m\}$. Let $W=\{w_1,\ldots,w_m\}$ be a set of codewords for the symbols in $\mathcal{X}$ with lengths $l_1, l_2, \ldots, l_m$.
Let $O_{i,j}$ be the subsets of $\mathcal{X}$ of possible symbols output by the source when transiting from state $S_i$ to state $S_j$, $O_{ij}$ being the empty set if transition from $S_i$ to $S_j$ is impossible.
If the code is uniquely decodable for the source $X$, then the matrix $\mathbf{Q}$ defined by
\begin{equation*}
\mathbf{Q}_{ij}=
\sum_{h \in O_{i,j}} D^{-l_h}
\end{equation*}
has spectral radius at most 1.
\end{thm}
\end{rem}

\begin{rem}[Shannon's insight]
\end{rem} 
An historical analysis reveals
that both McMillan's theorem and the proposed generalized one in the form of Theorem \ref{theo:modified-kraft-mealyform} are mathematically equivalent to a formulation obtained by Shannon already in \cite[Part I, Sec. 1]{shannon-1948} for the evaluation of the capacity of discrete noiseless channels.
In particular, in  \cite{shannon-1948} Shannon established that the capacity of an unconstrained noiseless channel with symbol durations $t_1,t_2,\ldots,t_m$ is given by the value $\log X_0$, where $X_0$ is the largest real solution of the difference equation 
\begin{equation*}\label{eq:shannon-sum}
X^{-t_1}+X^{-t_2}+\cdots+X^{-t_m}=1.
\end{equation*}
It is not difficult to show that McMillan's theorem is equivalent to the obvious statement that the capacity of a $D$-ary channel is at most $\log D$.

Furthermore, Shannon generalized the capacity formula to the case of noiseless finite state channels,  by stating the following \cite[Th. 1]{shannon-1948}:
\begin{quote}
\begin{thm}[Shannon, \cite{shannon-1948}]\label{theo:shannon-capacity2}
Let $b_{ij}^{(s)}$ be the duration of the $s^{\mbox{th}}$ symbol which is allowable in
state $i$ and leads to state $j$. Then the channel capacity $C$ is equal to $\log
W_0$ where $W_0$ is the largest real root of the determinant equation:
\begin{equation*}\label{eq:shannon-capacity2}
\left| \sum_s W^{-b_{ij}^{(s)}}-\delta_{ij}\right|=0.
\end{equation*}
\end{thm}
\end{quote}
As for the unconstrained case, it is possible to show that Theorem \ref{theo:modified-kraft-mealyform} is equivalent to the statement that every finite state $D$-ary channel has capacity at most $\log D$.

\bibliographystyle{IEEEbib}

\begin{thebibliography}{10}

\bibitem{verdu-tutorial}
S.~Verd\'u,
\newblock ``Fifty years of shannon theory,''
\newblock {\em IEEE Trans. on Inform. Theory}, vol. 44, no. 6, pp. 2057--2078,
  1998.

\bibitem{shannon-1948}
C.~E. Shannon,
\newblock ``A mathematical theory of communication,''
\newblock {\em Bell Sys. Tech. Journal}, vol. 27, pp. 379--423,623--656, 1948.

\bibitem{mcmillan-kraftineq}
B.~McMillan,
\newblock ``Two inequalities implied by unique decipherability,''
\newblock {\em IEEE Trans. Inform. Theory}, vol. IT-2, pp. 115--116, 1956.

\bibitem{kraft-kraftineq}
L.G. Kraft,
\newblock ``A device for quanitizing, grouping and coding amplitude modulated
  pulsese,''
\newblock M.S. thesis, Dept. of Electrical Engineering, MIT, Cambridge, Mass.,
  1949.

\bibitem{cover-thomas-book}
T.M. Cover and J.A. Thomas,
\newblock {\em Elements of Information Theory},
\newblock John Wiley, New York, 1990.

\bibitem{gallager-book}
R.G. Gallager,
\newblock {\em Information Theory and Reliable Communication},
\newblock Wiley, New York, 1968.

\bibitem{role-of-pattern}
A.~D. Wyner, J.~Ziv, and A.~J. Wyner,
\newblock ``On the role of pattern matching in information theory,''
\newblock {\em IEEE Trans. on Inform. Theory}, vol. 44, no. 6, pp. 2045--2056,
  1998.

\bibitem{Alon-onetoone}
N.~Alon and A.~Orlitsky,
\newblock ``A lower bound on the expected length of one-to-one codes,''
\newblock {\em IEEE Trans. on Inform. Theory}, vol. 40, pp. 1670--1772, 1994.

\bibitem{Wyner-onetoone}
A.~D. Wyner,
\newblock ``An upper bound on the entropy series,''
\newblock {\em IEEE Trans. on Inform. Theory}, vol. 20, pp. 176--181, 1972.

\bibitem{Blundo-one-to-one}
C.~Blundo and R.~De~Prisco,
\newblock ``New bounds on the expected length of one-to-one codes,''
\newblock {\em IEEE Trans. on Inform. Theory}, vol. 42, no. 1, pp. 246--250,
  1996.

\bibitem{Savari-onetoone}
S.~A. Savari and A.~Naheta,
\newblock ``Bounds on the expected cost of one-to-one codes,''
\newblock in {\em Proc. IEEE Intern. Symp. on Inform. Theory}, 2004, p.~92.

\bibitem{karush-1961}
J.~Karush,
\newblock ``A simple proof of an inequality of {M}c{M}illan,''
\newblock {\em IRE Trans. Inform. Theory}, vol. IT-7, pp. 118, 1961.

\bibitem{minc-book}
H.~Minc,
\newblock {\em Nonnegative Matrices},
\newblock Wiley, 1988.

\bibitem{sardinas-patterson-1953}
A.A. Sardinas and G.W. Patterson,
\newblock ``A necessary and sufficient condition for the unique decomposition
  of coded messages,''
\newblock in {\em IRE Convention Record, Part 8}, 1953, pp. 104--108.

\bibitem{DL_ITreport_2008}
M.~Dalai and R.~Leonardi,
\newblock ``On unique decodability, {McMillan}'s theorem and the expected
  length of codes,''
\newblock {\em University of Brescia, Technical Report R.T. 2008-01-58}, 2008.

\end{thebibliography}

\end{document}